\documentclass[twocolumn,prb]{revtex4}

\usepackage{graphicx}
\usepackage{bm}

\newcommand{\be}{\begin{equation}}
\newcommand{\ee}{\end{equation}}

\begin{document}

\title{Formation energy and interaction of point defects in two-dimensional
colloidal crystals}
\author{L.C. DaSilva$^{\dagger}$\cite{lder}, L. C\^andido$^{\ddagger}$, L. da F. Costa$^{\dagger}$
and Osvaldo N. Oliveira Jr$^{\dagger}$}
\affiliation{$^{\dagger}$Instituto de F\'isica de S\~ao Carlos, Universidade de 
S\~ao Paulo 13560-970, S\~ao Carlos, SP, Brazil \\
$^{\ddagger}$Instituto de F\'isica, Universidade Federal de Goi\'as, Campus II ,
74001-970 Goi\^ania, GO, Brazil}

\vspace*{3mm}

\begin{abstract}
The manipulation of individual colloidal particles using optical tweezers has
allowed vacancies to be created in two-dimensional (2d) colloidal crystals, with
unprecedented possibility of real-time monitoring the dynamics of such defects
(Nature {\bf 413}, 147 (2001)). In this Letter, we employ molecular
dynamics (MD) simulations to calculate the formation energy of single defects
and the binding energy between pairs of defects in a 2d colloidal crystal.
In the light of our results, experimental observations of vacancies could be
explained and then compared to simulation results for the interstitial defects.
We see a remarkable similarity between our results for a 2d colloidal 
crystal and the 2d Wigner crystal (Phys. Rev. Lett. {\bf 86}, 492 (2001)). The results
show that the formation energy to create a single interstitial is $12\% - 28\%$ lower 
than that of the vacancy. Because the pair binding energies of the defects are strongly
attractive for short distances, the ground state should correspond to bound pairs with
the interstitial bound pairs being the most probable.
\end{abstract}

\maketitle

\section{Introduction}
A revival of general interest in point defects in two-dimensional (2d)
systems has been fueled by demonstration that vacancies can be created
in 2d colloidal crystals through manipulation of individual particles
with optical tweezers\cite{Ling,Gast}. Because of the large size of
the colloidal particles, the structural and dynamical properties of
these point defects can be monitored with video microscopy
\cite{Ling2}. Interest in point defects in solids is widespread, 
including defects in ordinary materials as well as in quantum crystals 
such as $^4$He and Wigner crystals. In quantum crystals the point defects 
are believed to occur at finite concentrations at any nonzero temperature. 
There is also speculation that even at zero temperature point defects can 
exist, while at higher concentrations they may lead to a supersolid 
phase\cite{kin,cep,lc}. The role of point defects in melting in a 
two-dimensional system has been investigated theoretically\cite{fish,cock,jain,nel},
but an experimental investigation is usually hampered by the low concentrations
of defects, unless the temperature is close to the melting point\cite{mar}.

In this work, we report on the first accurate Molecular Dynamics
(MD) calculations of the formation energy of single point defects and
interaction of pair point defects in a 2d screened
Coulomb interaction colloidal system. The system is formed by colloidal 
particles, taken as identical spheres of radius $a$, suspended in a solvent 
(generally water) and confined between two parallel solid surfaces. When 
immersed in this solvent, the colloids acquire a large charge $Z$ due to 
dissociation of endgroups from their surface. Counterions thus generated ensure 
charge neutrality, forming a cloud around each colloid that makes the Coulomb 
interaction shorter. The colloids are free to move in 2d and interact through 
pair-wise Yukawa type potential\cite{hone,alex}. 
The Hamiltonian for this system is

\begin{equation}
H=\sum_{i}^{N}\frac{{\bf{p}_{i}}^2}{2m}+ \sum_{i<j}^{N}
\frac{(Z_{\lambda}e)^2}{\epsilon}\frac{e^{-\bf{r_{ij}}/\lambda}}{\bf{r_{ij}}}+NU_{B},  
\end{equation}
where the first term in the right is the kinetic energies, the
second is the screened Coulomb colloid-colloid interaction and the
third term is the interaction between the colloids and  the neutralizing 
background of positive charges\cite{pet} $U_{B} =-2\pi b\lambda/a_{0}^2$, 
where $b=(2/\sqrt{3})^{1/2}a_{0}$ is the lattice space and $a_{0}$
is the average separation between colloids (this is defined for the triangular 
lattice with vector translation $(b,0)$ and $(b/2,b\sqrt{3}/2)$ with 
$a_{0}=1/\sqrt{\rho}$ where $\rho$ is the 2d colloid number density),
 $\lambda$ is the screening length, $\epsilon$ is the dielectric constant of
 the medium and $Z_{\lambda}=Z^{*}f(a/\lambda)$, where $Z^{*}$ is the normalized
charge and $f(x)=\sinh(x)/x$ is a function that describes the effect of the
nonzero radius $a$ of the colloidal particles.  The energy, length,
temperature and time are in units of $E_{0}=(Z_{\lambda}e)^{2}/\epsilon \sigma$, 
where $\sigma = 1.1 \mu$m (typical lattice space for experimental systems),
$T_{0}=E_{0}/\kappa_{B}$ ($\kappa_B$ Boltzmann constant) and
$t_{0}=(E_{0}/m\sigma^2)^{-1/2}$, respectively.

\section{Computational details}

The system with point defects (vacancy or interstitial) is
modeled by removing (adding) a colloidal particle from (to) the most
stable 2d lattice, which is the triangular lattice. We place the
defect at the center of the simulation box to avoid complications
arising from lattice relaxation: a single 6 coordinated vacancy at the
central lattice site or a threefold centered interstitial colloid in
one of the central triangular unit cells. However, no constraint
exists which would restrict the center of each colloid to lie within
its own Wigner-Seitz cell, i.e. to make the lattice relaxation 
locally. This means that the defects are free to move around and can change
their symmetry during the system evolution in thermodynamic
equilibrium. 

In order to calculate the energy needed to create one single defect we
perform two independent simulations at the same density and
temperature: a simulation for the perfect and a simulation for the
defective system. For the latter, after introducing the defect, we
rescale the dimensions of the simulation box by a factor $f =
\sqrt{n_{d}/n_{p}}$ to reset the system to the original density, with
$n_{d}$ and $n_{p}$ being the number of colloids for the defective and the
perfect system, respectively. This is performed to circumvent the need
of correcting the energy due to density change caused by inclusion of
the defect. The difference between the energies of the defective and
the perfect systems is the energy needed to create the
defect. Formally, we can define the number of defects $N_{def}$ as the
number of colloids minus the number of lattice sites.  Therefore, the
formation energy of $N_{def}$ defects in a crystal with $N$ lattice
sites is

\begin{equation}
\Delta E_{def}=[\tilde{E}(N+N_{def})-\tilde{E}(N)](N+N_{def}),
\end{equation}
where $\tilde{E}(n)$ is the energy per colloid for a system containing
$n$ colloids. For a monovacancy or an interstitial defect,
$N_{def}=N_{v}=-1$ and $N_{def}=N_{int}=+1$, respectively. 

MD calculations of the formation energy of point defects were
performed for several system sizes $n=29$, $30$, $31$, $129$, $130$,
$131$, $269$, $270$, $271$, $479$, $480$ and $481$ colloids, which
allows one to study finite-size effects at different densities. Larger
system sizes would be unpractical computationally because high
accuracy is needed in order to obtain the energy differences. 

We used colloids with radius $a\sim 0.18$ $\mu$m,
screening length $\kappa^{-1}\sim 0.39$ $\mu$m, charge $\sim 1650e$
and density varying from $\rho=0.402$ to $1.804$ $(\mu$m$)^{-2}$,
corresponding to typical experimental data \cite{Ling}.
To give an idea of the energies and temperatures involved in
the calculations, for instance at $\rho=0.954(\mu$m$)^{-2}$ the units are
$E_{0}=9.4\times10^{-18}$ Joules and $T_{0}=6.81\times10^{5}$K.

The initial positions for the colloids are the sites of a triangular lattice
accommodated in a rectangular box with periodic boundary conditions to
eliminate surface effects. The simulations were performed within the
canonical ensemble keeping a constant system temperature using the Berendsen's
thermostat with coupling parameter to an external bath $\tau_{T}$
ranging from $0.01$ to $0.1$ \cite{berend}. The evolution of Newton's
equation of motion is obtained with the four-order predictor-corrector
algorithm. The time step varies from $2.5\times 10^{-2}$ to
$5.0\times 10^{-3}$ $t_{0}$ since it has some scale
dependence on the colloid density. Thermodynamic equilibrium was
assumed to be achieved during the first 50000 time steps, after which
the physical quantities were obtained by averaging over 700 blocks of
10000 time steps.

\section{Results and discussion}

In the simulation, the formation energy of the vacancy and interstitial defects
depends on the number of particles, i.e. size of the simulation
box. To eliminate the size dependence of the formation energy we extrapolate 
the simulation results to the thermodynamic limit by using the following formula

\begin{equation}
\Delta E_{def}^{N}(\rho)=\Delta E^{\infty}_{def}(\rho)+\frac{c(\rho)}{N},
\end{equation}
where the density-dependent parameters $\Delta E^{\infty}_{def}(\rho)$
and $c(\rho)$ are determined by a linear least-squares fit to the MD
calculations at different $N$.  The MD formation energies for the
point defects and $\chi^2$ of the fitted data as a function of $\rho$,
at a fixed temperature and screening length, are shown in table I. The
$\chi^2$ of the fitted data is particularly good, indicating that the
size dependence is well described by Eq. (3).  The formation energies
for interstitial defects are consistently lower, from $12\% - 28\%$,
than those for vacancies in the range of densities studied. Therefore,
an interstitial defect is more stable and more likely to be created.

The formation energy of the defects is practically temperature-independent
in the temperature range $3.\times10^{-5}T_{0} < T < 5.\times10^{-2}T_{0}$
below the melting point, as shown in Fig. 1, where the formation
energy of the interstitials is smaller than that of the vacancies.
However, when the kinetic energy starts to compete with the potential energy
close to the melting point ($T > 5.0\times10^{-2}T_{0}$), the formation energy of the
defects becomes too noisy due to large fluctuations in the total energy and
approaches zero within the statistical error. This suggests that single point
defects may be easily created thermally in this temperature range
and could play an important role in the melting mechanisms of 2d
colloidal crystals.
   
\begin{figure}
\includegraphics[width=6cm,height=6cm]{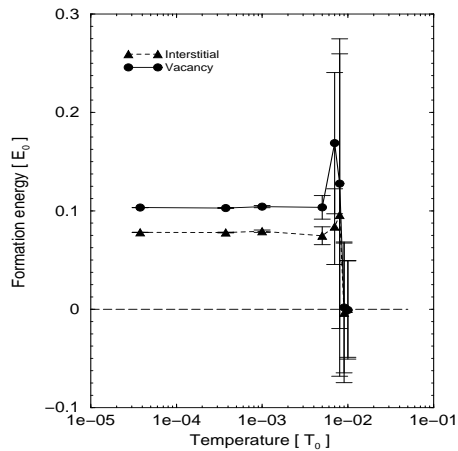}
\caption{Formation energy of a vacancy and interstitial defect as a function
of temperature at $\rho=0.954$ $(\mu$m$)^{-2}$. Computed for systems of 120 lattice sites.}
\end{figure}

\begin{table*}
\caption{Formation energy in reduced units obtained from MD
simulations for a vacancy (left) and interstitial defect (right) for
various system sizes denoted by $N$ (number of lattice sites) and at various values of
$\rho$ in $(\mu$m$)^{-2}$. Quantities in parentheses are the estimated error in the last
decimal place. Also given are the energies for an infinite system
($\Delta E^{\infty}_{Vac}$, $\Delta E^{\infty}_{Int}$), c($\rho$) and
$\chi^2$ fitting parameters. The system temperature and the screening
length are  $T=3.74\times 10^{-4}$$T_{0}$ and $\kappa^{-1}=0.39$ $\mu$ ,
respectively.}

\begin{tabular}{ccccc||ccccc} \hline \hline
\multicolumn{5}{c||}{Vacancy} & \multicolumn{5}{c}{Interstitial} \\ 
  & $\rho=0.402$ & $\rho=0.589$ & $\rho=0.954$ & $\rho=1.804$ &
  & $\rho=0.402$ & $\rho=0.589$ & $\rho=0.954$ & $\rho=1.804$ \\ \hline 
 $N=30$   & 0.0261(5)    & 0.0543(4) &          &           &
 $N=30$   & 0.0238(5)    & 0.0430(4) &          &           \\ 
 $N=120$  & 0.029(1)     & 0.0550(4) &0.1029(4) & 0.1933(4) & 
 $N=120$  & 0.024(1)     & 0.0435(6) &0.0781(5) & 0.1425(3) \\ 
 $N=270$  & 0.029(2)     & 0.055(1)  &0.1032(4) & 0.1958(7) &
 $N=270$  & 0.024(2)     & 0.043(3)  &0.0783(6) & 0.1426(4)  \\ 
 $N=480$  & 0.029(3)     & 0.055(1)  &0.1032(9) & 0.1958(6) & 
 $N=480$  & 0.024(2)     & 0.043(6)  &0.0785(5) & 0.1428(5) \\  
 $\Delta E^{\infty}_{Vac}$& 0.029(1)  &0.0551(4) & 0.1033(4) & 0.1969(7)&
 $\Delta E^{\infty}_{Int}$& 0.024(1)  & 0.0436(7)& 0.0785(6) & 0.1428(5) \\  
 c($\rho$)                & -0.10(3)  &-0.02(2)  & -0.05(8)  & -0.4(1)  &
 c($\rho$)                & -0.00(3)  &-0.01(3)  & -0.0(1)   &  -0.03(8) \\  
 $\chi^2$                 &  0.086    & 0.032    &  0.023    &  0.6    &
 $\chi^2$                 &  0.00056  & 0.046    &  0.018    &  0.049  \\ \hline\hline  
\end{tabular}
\end{table*}

We now investigate the possible defect topologies as the system
evolves in thermodynamic equilibrium, which is particularly relevant
for defining the dynamics of the defects, as observed
experimentally\cite{Ling} and recently corroborated by results from
Brownian simulation methods\cite{Lib}. According to our simulations,
the initial configurations for the threefold centered interstitial
and the sixfold vacancy were found to relax into a configuration of
lower symmetry, in agreement with simulation and experimental
observations. However, the aforementioned observations concerning the 
dynamics of defects focus only on the symmetry and topology of the
defect. In the following we discuss the dynamics of the defects in
topological and energetic terms.

As stated in Ref. \cite{Ling2}, at finite temperatures this many body
system vibrates around every local energy minimum due to thermal
fluctuations. If the energy differences between distinct local minima
are small, the system can get enough energy to move to a
nearby local minimum. As long as the system remains around a local
energy minimum, the distortions in the lattice are elastic and the
topological arrangement of the particles does not change. This allows
us to calculate the system energy for each topology. For such a calculation, 
the defect must be tracked after the system has reached thermodynamic 
equilibrium. For that, we developed a code to perform a dynamical check 
of a list of neighbors of each colloid in a triangular lattice. This list is
created (updated) by counting the sides of the polygons in the Voronoi
\cite{voro} construction at each time step run. A defect is
characterized by the presence of miscoordinated particles,
i.e. particles whose number of neighbors is different from 6. Once the
current topology of the defect is identified, the corresponding energy
is recorded. We also calculated the time the defect remains
in a given topology and the number of transitions each defect performs  
between different topological configurations. 

Table II summarizes our findings for the dynamics of the defects. The
topologies for the vacancy are: crushed vacancy ($V_{2}$); symmetric
vacancy ($V_{3}$); split vacancy ($SV$); which were observed
experimentally \cite{Ling}, and another one, a fourfold symmetric
excited configuration ($V^{'}_{4}$), only observed recently
\cite{Lib}.  For interstitial defects the topologies are: threefold
symmetric interstitial $(I_{3})$; twofold symmetric
interstitial $(I_{2})$; disjoint twofold symmetric
interstitial $(I_{2d})$; and a fourfold symmetric excited
interstitial ($I^{'}_{4}$). The transition matrix (a
stochastic matrix) in the upper part of Table II indicates that for a
sufficiently long time, both vacancy and interstitial defects adopt
the possible topologies many times. Each row of this matrix gives the
probability of transition from a state (topology) to another one. 
As pointed out in Ref. \cite{Lib}, the topologies $V_{2}$, $SV$, $I_{2}$ 
and $I_{2d}$ have a preferential diffusion direction, with motion being a 
random walk along the main crystalline directions. In contrast, the $V_{3}$, 
$V^{'}_{4}$, $I_{3}$ and $I^{'}_{4}$ have no preferential direction and may 
act in switching the direction of motion. Therefore, the diffusion process of 
our system, for a long time run, is isotropic. However, we observed in subsidiary
simulations (results not shown here) that for short runs there is a
very small number of transitions to topological configurations
responsible for changing the direction of motion ($V_{3}$, $V^{'}_{4}$,
$I_{3}$ and $I^{'}_{4}$), especially for the vacancy. As a result,
diffusion becomes one-dimensional, consistent with experimental data
and simulation results\cite{Ling,Lib}. 

At the bottom of Table II, we show the times ($\tilde{t}_{s}$) for
the defects in each topological configuration in equilibrium, as well as 
the formation energy $\Delta E$ for each topological configuration. 
The formation energy for the various topologies are very close to each 
other, being almost within the statistical error. 
However, one infers that the lowest formation energy for both vacancy and 
interstitial defects correspond to the configurations $S_{V}$ and $I_{3}$, where 
in average the defects remain most of the time ($\tilde{t}_{s}$). This result
should be expected as these configurations are the most likely to be
thermally activated since they need the smaller amount of energy to be
created. On the other hand, the defect tends to spend a very short time in
$V^{'}_{4}$ and $I^{'}_{4}$ configurations because they have the
highest formation energies. Note that $I_{2d}$ topology is just a
transient variation of $I_{2}$ topology.
A direct comparison of the formation energies for each topological configuration
and its lifetime can lead to wrong interpretation of the results of Table II. 
Although the defects formation energies between such topological configurations are
similar, the average time of the defect in each topological configuration is quite
different due to differences in the energy barriers.
For instance, though the formation energy of $V_2$ is slightly smaller than
that of $V_3$, the lifetime of $V_2$ is much larger. The reason is that the net
transition probability from $V_3$ to $SV$ is much larger than that from $V_2$ to
$SV$ (about, $36\%$, see transition matrix), which indicates a smaller energy barrier between
$V_3$ and $SV$ topologies making $V_3$ less stable than $V_2$. 
One can make also the same discussion for the interstitials defect (right hand
side of the table II).

Far from the melting point, an increase in temperature causes the
number of transitions to increase, including the low-frequency
transitions between different topological configurations. Close to the
melting transition, the formation energy for different topological
configurations is difficult to calculate owing to the large thermal
fluctuations, with the differences in energy lying within statistical
error. 

\begin{table*}

\caption{The upper part is the normalized transition matrix  
for different topological configurations of the defects, during $\sim
7.0\times10^{6}$ MD steps after the system has reached equilibrium. In
the bottom are displayed the time spent in each
topological configuration ($\tilde{t}_s$) and the formation energy in reduced units
of the defects, $\Delta E$, for the 2d colloidal crystal at $\rho=0.954$ $(\mu$m$)^{-2}$.
Quantities in parentheses are the estimated errors in the last decimal place.
The system temperature and screening length are $T=1.0\times10^{-3}$ $T_{0}$ and
$\kappa^{-1} =0.39$ $\mu$m,
respectively.}

\begin{tabular}{ccccc||ccccc} \hline \hline
\multicolumn{5}{c||}{Vacancy} & \multicolumn{5}{c}{Interstitial} \\ 
           & $V_{2}$ &$V_{3}$ & $SV$   & $V^{'}_{4}$ &  
           & $I_{2}$ &$I_{2d}$& $I_{3}$& $I^{'}_{4}$   \\ \hline  
$V_{2}$ &0.9987098      & 0.0000154     & 0.0012747  & 0       & 
   $I_{2}$ & 0.9977724       &  0.0000890     & 0.0021340   & 0.0000046    \\ 
   $V_{3}$&  0.0000175      &0.9995009        & 0.0004666    &0.0000150  &
   $I_{2d}$& 0.0051196      &0.9947921        &0.0000883      &0  \\
   $SV   $ & 0.0009514          &0.0000255         &0.9990231     &0.    &
   $I_{3}$ & 0.0007948         &0.0000007116  &0.9989618     &0.0002426  \\
   $V^{'}_{4}$ &0.                         &0.0048446         & 0                     &0.9951554  & 
   $I^{'}_{4}$ &0.0000450           &0.0000064         &0.0054687     &0.9944798   \\ \hline
 $\tilde{t}_{s}$      &0.4068799&0.0420110&0.5509791&0.0001300 &  
 $\tilde{t}_{s}$      &0.2618633&0.0046109&0.7024319&0.0310939\\
 $\Delta E $&0.1044(6)&0.1046(6)&0.1040(6)&0.1086(6) &  
 $\Delta E $&0.0798(6)&0.0799(6)&0.0790(6)&0.0829(6) \\ \hline\hline
\end{tabular}
\end{table*}

\begin{figure}
\includegraphics[width=6cm,height=6cm]{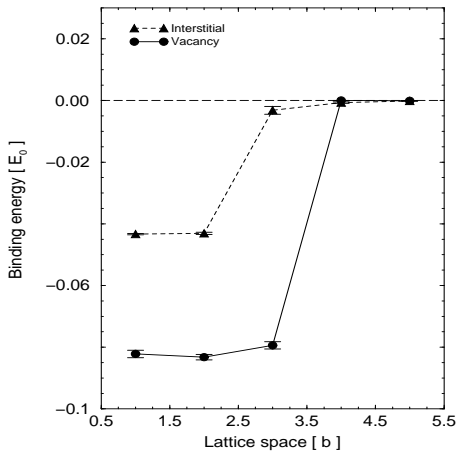}
\caption{Binding energy of a pair defect of vacancy-vacancy and interstitial-interstitial
as a function of lattice spaces at $\rho=0.954$ $(\mu$m$)^{-2}$. Computed for systems
of 120 lattices sites.}
\end{figure}

We also studied the interaction of a pair of defects as a function of
the lattice separation. The difference between the energies to create
two single defects separated by a given number of lattice constants
and the energy to create two isolated single defects is the binding
energy between two single defects, defined as follows
    
\begin{equation}
E_{int}=\Delta E_{def}(N_{def}=2)-2\Delta E_{def}(N_{def}=1)
\end{equation} 

where the first term in the right $\Delta E_{def}(N_{def}=2)$
corresponds to the energy to create two single defects separated by
some lattice constants, while the last term $2\Delta
E_{def}(N_{def}=1)$ is the energy to create two isolated single
defects.

Fig. 2 shows the binding energy vs. lattice separation.  In order to
enforce accuracy in positioning the pairs of defects, we slow down the
movement of the colloids by decreasing the temperature to $\sim
10^{-10}$ $T_{0}$. It was enough to hold the two defects in a fixed
separation. Both the pairs of vacancy-vacancy and interstitial-interstitial
are strongly attractive at short distances, with attraction going to zero for
distances greater than three lattice spaces. For short distances
the binding energies for the defects are $\sim -0.08$ and $\sim
-0.04$ for the vacancy and interstitial defect, respectively, being
therefore higher than the value expected at the melting temperature at
density $\rho=0.954$ $(\mu$m$)^{-2}$, which is $\sim 0.01$ according to Fig. 1.
We have also (not shown here) calculated the binding energy as a
function of the density for just one lattice space separation between
defects, which resulted always attractive in the range of densities
$\rho=0.402-1.804$ $(\mu$m$)^{-2}$. Since any attraction should suffice to permit
recombination, our results suggest that the ground state energy of 
the 2d colloidal crystal may be dominated by pair binding of defects. 
Furthermore, as interstitial defects have the lowest excitation energy,
we expect that the ground state of point defects should involve mainly interstitial
pairs. Though the mechanisms of pair binding of point defects in solids are still not
fully understood, we note that our results are similar to those for interstitial
defects in quantum crystals, such as vortex crystals\cite{jain} and 2d Wigner
crystals\cite{lc}, in which the defect pair is not sufficiently strong to yield
a supersolid phase.
For experimentalists working in the 2d colloidal system using
video microscopy, we believe the results on the pair binding of defects
provide an experimental challenge to observe formation of pair point defects near the 2d
colloidal crystal melting.   

Before concluding, we comment on the entropic factors, which are embedded in the
calculations of the formation energies.
Such factors could be neglected well below the melting temperature, as they are
much less important than the energy terms. Close to the melting temperature,
entropic factors are likely to affect the numerical values of the formation energies,
but the qualitative features should be preserved. Indeed, this expectation appears
to be fulfilled in the experimental system represented by the 2d colloidal crystal,
since our simulations could explain the experimental observations. 

In conclusion, we have shown quantitatively the effects from point
defects (vacancy and interstitial) in 2d colloidal crystals, in which the energy
to create a single interstitial defect is $12\%- 28\%$ lower than to create a single
vacancy.  The formation energies of these defects
go to zero near the melting point, i.e. point defects can be easily
created thermally and should play a crucial role in the melting
mechanisms for such crystals. We also confirmed previous results that
the interstitial defects are more mobile than vacancies, and provided an
explanation based on energy calculations. Finally, we found that the
interaction between defects is strongly attractive, and consequently
most defects will exist as bound pairs. 
We believe that our results may have important bearing on experimental
works involving interfaces and solid surfaces.

\begin{acknowledgments}
This research was supported by CNPq, CAPES, FAPESP and FUNAPE-UFG.
Luciano da F. Costa thanks FAPESP (05/00587-5) and CNPq (308231/03-1)
for financial support. The authors thank professor G.-Q. Hai for 
useful discussions.
\end{acknowledgments}

\end{document}